\documentclass[prl,twocolumn,showpacs]{revtex4}

\usepackage{graphicx}
\usepackage{bm}

\begin{document}
\title{Full Current Statistics in the Regime of Weak Coulomb Interaction}
\author{D.~A. Bagrets and Yu.~V. Nazarov}
\address{Department of Applied Physics and 
Delft Institute of Microelectronics
and Submicrontechnology, \\
Delft University of Technology, Lorentzweg 1, 2628 CJ Delft, The Netherlands
}
\date{\today}

\begin{abstract}
We evaluate the full  statistics 
of the current via a Coulomb island that is  
strongly coupled to the leads. This strong coupling weakens
Coulomb interaction. We show that in this case
the effects of the interaction can be incorporated 
into the renormalization of 
transmission eigenvalues of the scatterers that 
connect the island and 
the leads. We evaluate the Coulomb blockade gap in the
current-voltage characteristics, 
the value of the gap being exponentially 
suppressed as compared 
to the classical charging energy of the island. 
\end{abstract}

\maketitle

The electric charge is quantized. There are two important
manifestations of this fact in electron transport on
mesoscopic scale.

The first manifestation is a shot noise~\cite{BlanterReview}. 
The quantum mechanics modifies the shot noise
in an arbitrary mesoscopic conductor 
with respect to its classical value $S=2eI$. 
It also changes the full
statistics of charge transfer via a mesoscopic scatter, 
so that the statistics is not the simple Poissonian one~\cite{Levitov}. 

The second manifestation of electron charge quantization
is the Coulomb blockade phenomenon. 
The Coulomb blockade is most strong provided 
the conductance $G$ of a mesoscopic system is much smaller than
the conductance quantum $G_Q = e^2/2\pi\hbar$~\cite{IngoldNaz}. 
The statistics of charge
transfer in this strongly interacting case 
is that of a classical stochastic Markov process ~\cite{CB_FCS}. 
On the contrary, quantum mechanics is of importance for
Coulomb blockade in the opposite limit 
$G\gg G_Q$ where Coulomb 
interaction is weakened. 

The charge quantization manifests itself in both ways in 
full current statistics under conditions of weak Coulomb interaction,
this provides the motivation for the present study.
 
It has been understood that the charge quantization
survives even in the limit  $G\gg G_Q$
~\cite{Panyukov, Grabert, Zwerger, Matveev, Flensberg, Nazarov}.
The ground state energy of the Coulomb island was shown to
retain the periodic dependence on the induced "off-set" charge $q$, 
thus indicating the Coulomb blockade. 
However, due to large quantum 
fluctuations of charge in the island, 
the effective charging energy $\widetilde E_C$ (defined 
as the periodic $q$-dependent part of free energy) 
is exponentially
suppressed as compared to the classical charging energy $E_C = e^2/2C$. 
It is important that the weak charge quantization 
persists not only for tunnel
junctions ~\cite{Panyukov, Grabert, Zwerger} connecting the island and the 
leads. 
The connectors can be arbitrary mesoscopic scatterrers~\cite{Nazarov}. 
The quantization completely vanishes only for constrictions
with perfectly transmitting channels~\cite{Matveev, Flensberg}. 

This research was concentrated on the equilibrium properties, 
an alternative is to study transport. 
Recent studies link the short noise in the conductor to 
the interaction correction to the conductance. \cite{corrnoise}
The interaction correction to full current statistics was
analyzed in \cite{Kindermann} for a scatterer embedded in the 
electromagnetic environment.
The relation between interaction correction to the
conductance and the formation of Coulomb blockade in an
island was investigated in \cite{GolubevZaikin}.

In this Letter we evaluate the full statistics 
of the current via the Coulomb island defined by several
mesoscopic scatterers assuming weak Coulomb interaction, $G\gg G_Q$.
The results can be summarized as follows. At energy scale 
$E \ll g_0 E_c$, $g_0=G_0/G_Q$ being 
the dimensionless conductance of the system in the absence of 
interaction, the dominant effect 
of Coulomb interaction is the energy-dependent renormalization
of the transmission eigenvalues $T_n^{[k]}$ of the 
mesoscopic scatterers labeled by $k$, 
\begin{equation}
 \frac{d\, T_n^{[k]}}{d\ln E} = 
\frac{2\, T_n^{[k]}(1- T_n^{[k]})}{\sum_{n,k}  T_n^{[k]}}.
\label{RGEquation}
\end{equation} 
The renormalization of a similar form was previously 
obtained in \cite{Glazman}
for a scatteter in the weakly interacting 1d gas and in \cite{Kindermann}
for a single multi-channel scatterer shunted by an external impedance. 
We thus prove this simple relation for a Coulomb island.
The full current statistics is readily obtained from the energy-dependent
$T_n^{[k]}$ with using non-interacting 
scattering theory approach of \cite{NazBag,Levitov}.
This gives the voltage dependence of conductance, shot noise and 
all higher cumulants of charge transfer. In contrast to the case of
a single scatterer, the renormalization of all transmission eigenvalues
may break down at finite energy 
$\widetilde E_C \propto   g_0 E_C e^{-\alpha g_0}$, 
$\alpha$ being a numerical factor depending on the details
of the initial transmission distribution.
Remarkably, $\widetilde E_C$ coincides 
with the equilibrium effective charging energy evaluated with
instanton technique.\cite{Nazarov} However, the renormalization
stops at the effective Thouless energy  
$E_{\rm Th} \sim G(E) \delta/G_Q$, 
$\delta$ being mean level spacing in the island. 
This gives rise to 
{\it two} distinct 
scenarios at low enegy.
If $g_0 > \alpha^{-1}\ln(E_C/\delta)$, Coulomb blockade does not
occur with zero-bias conductance being saturated 
at the value $G(E_{\rm Th}) \gg G_Q$.
Alternatively, 
$G(0)\approx 0$ and $\widetilde E_C$ defines the Coulomb gap.  
 
\begin{figure}[t]
\begin{center}
\includegraphics[width=2in]{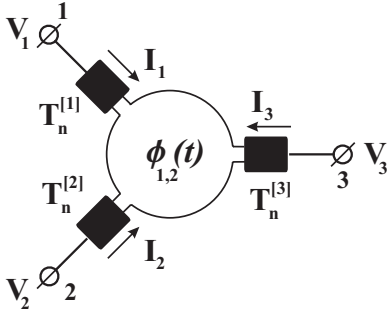}
\caption{Multi-terminal (M=3) Coulomb island. }
\end{center}
\end{figure}

Let us give the details of the model in use. 
The Coulomb island is characterized by two parameters:
charging energy $E_C$ and the mean level spacing $\delta$, $E_C \gg \delta$. 
The island is connected to $M \ge 2$ external leads
by means of $M$ arbitrary mesoscopic scatterers. (Fig. 1)  
Each scatterer, or connector, $i$ is characterized by the set
of transmission eigenvalues $T_n^{[i]}$.  
We assume that the island is strongly coupled to the leads,
$g_0 = \sum_{n,m} T_n^{[i]} \gg 1$.
Our goal is to evaluate the 
cumulant generating function (CGF) $S([\chi_i]))$.
The Fourier transform of $\exp(-S)$ 
with respect to counting fields $\chi_i$
gives the probability $P(\{N_i\})$ 
for $N_i\gg 1$ electrons to be transferred to the terminal
$i$ during time interval $t_0$. (See~\cite{Levitov}) 
The derivatives of $S$ give the average value of currents,
shot-noise correlations and higher order moments of charge transfer.

To evaluate the CGF of the Coulomb island, we have extended
the semiclassical approach for the FCS of the non-interacting
electrons~\cite{NazBag}. 
To account for Coulomb interactions, we introduce
a dynamical phase variable $\varphi(t)$~\cite{Schoen} 
that results from the Hubbard-Stratonovich transform  
of the charging energy. Its time derivative, $\dot\varphi(t)/e$,
presents the fluctuating electrostatic potential of the island.
The CGF $S(\{\chi_i\})$ can be then represented in the form of a 
real-time path integral over the fields $\varphi_{1,\,2}(t)$ residing
at two branches of the Keldysh contour
\begin{eqnarray}
&&\exp(-S\{\chi_i\}) = \int D\varphi_{1,2}(t)\exp\Big\{
\frac{i}{2}E_C^{-1}\int\limits_{-\infty}^{+\infty} d\,t
(\dot\varphi_1^2-\dot\varphi_2^2) \nonumber \\
&& - \sum_k S^{[k]}_{\rm con}\bigr(\{\hat G,\hat G_k^\chi\}\bigl) \,-\,
 i\pi\delta^{-1}{\rm Tr}\{(i\partial_t - \dot\Phi)\hat G\} 
\Big\} \label{Model}  
\end{eqnarray}
Here  
$
\hat\Phi = \left(
\begin{array}{cc}
  \phi_1(t) & 0 \\
   0  &  \phi_2(t)
\end{array}
\right)
$ is the matrix in Keldysh space, $2\times 2$ matrix $\hat G(t_1,t_2)$
presents  the electron Green function in the island
that implicitly depends
on $\varphi_{1,2}(t)$. 
The trace operation includes the summation
over Keldysh indices and the integration in time. 
The contribution of each connector $S^{[k]}_{\rm con}$ 
has a form found in the circuit 
theory~\cite{NazBag,Wolfgang}
\begin{equation}
 S_{\rm con}^{[k]} = -\frac{1}{2}\sum_{n} {\rm Tr} \ln\left[1+ \frac{1}{4}T_n^{[k]}
(\{\hat G,\hat G_k^\chi\} - 2)\right] 
\label{Sk}
\end{equation} 
$\{\hat G,\hat G_k^\chi\}$ denoting the
anticommutator of the Green functions with 
respect to both Keldysh and time
indices. The Green functions in the leads $\hat G_k(\chi)$  
are obtained by $\chi$-dependent gauge
transformation ~\cite{Wolfgang} 
of the equilibrium Green functions in the reservoir $k$, 
$\hat G_k^{[0]}$,
$
\hat G_k^\chi(\epsilon) = \exp (i \chi_k \bar \tau_3/2) \hat G^{(0)}_k(\epsilon) 
\exp (-i \chi_k \bar \tau_3/2)
\label{boundary}
$,
where $\hat G_k^{[0]}$ are given by
$
\bar G_k^{[0]} = \left(
\begin{array}{cc}
 1-2f_k & -2 f_k \\
 -2(1-f_k) & 2f_k-1
\end{array} \right) 
$.
Here  $f_k(\varepsilon)$ 
presents the electron distribution function in the $k$-th reservoir.
The expression (\ref{Sk})
is valid under assumption of instantaneous electron transfer
via a connector, this corresponds to energy independent $T^{[k]}_n$. 

In order to find $\hat G(t_1,t_2)$ at given $\varphi_{1,2}(t)$,
we minimize the action with respect to all $\hat G(t_1,t_2)$ 
subject to the constrain 
$\hat G \circ \hat G = \delta(t_1-t_2)$.
This yields the saddle point equation for  
$\hat G(t_1,t_2)$:
\begin{equation}
 \sum_{n,\,k}
  \frac{T^{[\,k]}_n [\hat G_k^\chi,\hat G ]}{
    4+T^{[\,k]}_n\left(\{\hat G_k^\chi, \hat G\} - 2 \right)} = 
  i\pi\delta^{-1}[\,i\partial_t - \dot\Phi, \hat G ] \label{Saddle}
\end{equation}
where $[..\,,..]$ denotes the commutator in the Keldysh-time space. 
This relation
expresses $\hat G(t_1,t_2)\equiv \hat G(t_1,t_2; 
[\varphi_{1,2}(t)])$  
via the reservoir Green functions $\hat G^{[k]}$. 
This circuit theory relation is similar to obtained in \cite{NazBag}.
It disregards the mesoscopic fluctuations, since those
lead to
corrections of the order of $\sim 1/g_0$ at all energies, 
whereas the interaction corrections are of the order of $\sim 1/g_0 \ln(E)$
tending to diverge at small energies. 
If $\varphi_{1,2}(t)=0$, 
Eq.~(\ref{Saddle}) separates in energy representation
and coincides with that of Ref. \cite{NazBag}.

This sets the model.
We start the analysis of the model with
perturbation theory in  $\varphi_{1,2}$
around the semiclassical saddle point $\hat G(t_1,t_2)=\hat G_0$,
$\varphi_{1,2}(t)=0$. 
The phase fluctuations are
small, $\delta \varphi^2 \sim 1/g_0$, so we 
keep only  quadratic 
terms to the action ~(\ref{Model}).
The resulting Gaussian path integral over $\varphi_{1,2}$ 
can be readily done.
This procedure is equivalent to the summation
of all one-loop diagrams of the conventional perturbation theory, i.e.  
to the  "random-phase approximation" (RPA).

For the rest, we restrict ourselves
 to the most interesting 
low voltage/temperature limit, $\max\{eV,kT\}\ll g_0 E_C$.
In this limit, we evaluate the interaction correction to the CGF
with the logarithmic accuracy.
It reads
\begin{eqnarray}
&&\Delta S_\chi = \frac{t_0}{g_0}\ln\left(\frac{g_0 E_C}{\max\{eV,kT\}}\right) \times 
\label{S2} \\
&&\int \frac{d\varepsilon}{2\pi} \sum_{n,\,k}
\frac{2T^{[\,k]}_n(1-T^{[\,k]}_n)\bigl(\{\hat G_k^\chi,\hat G_0 \}-2\bigr)}{
    4+T^{[\,k]}_n\bigl(\{\hat G_k^\chi, \hat G_0\} - 2 \bigr)} \nonumber
\end{eqnarray}
provided $\max\{eV,kT\} > E_{\rm Th}$, 
where $E_{\rm Th}=g_0\delta$ is the Thouless energy of the island. 
In the opposite case,
$\max\{eV,kT\}< E_{\rm Th}$, 
the voltage/temperature should be replaced
with $E_{\rm Th}$.
Note, that the correction (\ref{S2}) is contributed by only 
virtual inelastic processes that change the probabilities
of real elastic scatterings.

For simplicity, we consider the shot-noise limit
$eV\gg kT$ only.  
Then the magnitude of the correction  
shall be compared with the zero-order CGF $S^{[\,0]}_\chi\sim t_0 eV g_0$. 
This implies that the perturbative RPA result~(\ref{S2}) 
is applicable only if 
$\displaystyle g_0^{-1}\ln\left({g_0 E_C}/{eV}\right)\ll 1$.
At lower voltages $\Delta S_\chi$ 
logarithmically diverges. This indicates that we should
proceed with a renormalization group (RG) analysis.
   
We perform the RG analysis of the action~(\ref{Model}) 
in the one-loop approximation, this corresponds to 
the "poor's man scaling". This is justified by $g \gg 1$.
We follow the conventional procedure \cite{renormalization}
and decompose $\varphi(t)$ onto the fast 
$\varphi_f$ and slow parts $\varphi_s$. 
On each step of RG procedure we
eliminate the fast degrees of freedom in
the energy range $E-\delta E<\omega<E$ 
to obtain new action $S_{E -\delta E} [\varphi_s]$,
$E$ being the current ultraviolet cutoff.
Our key result is that 
the change in the action at each step of RG procedure can be
presented as a change 
of transmission eigenvalues
$T_n^{[k]}$.

Therefore, the RG equations can be written directly for
transmission eigenvalues and take a simple form (\ref{RGEquation}).
The equations are to be solved 
with initial conditions at the upper cutoff energy 
$E = g_0 E_C$, those are given by "bare" transmission
eigenvalues $T_n^{[k]}(E = g_0 E_C) =T_n^{[k]}$. 
The RG equations resemble 
those for the transmission coefficient 
for a scatterer in the weakly interacting one-dimensional electron 
gas~\cite{Glazman} and for a single multi-channel scatterer in the
electromagnetic environment \cite{Kindermann}.
The effective impedance $Z$ is just replaced by
inverse conductance of the island to all
reservoirs, $G(E)= G_Q\sum_{n,k} T_n^{[k]}(E)$.
The important difference is that this conductance is itself
subject to renormalization. 
The difference becomes most evident in the case when 
all  contacts are tunnel junctions, $T_n^{[k]}\ll 1$. In this
case, one can sum up over $k,n$ in 
Eqs. (\ref{RGEquation}) to obtain the RG for the
conductance only : $dG/d\ln E = 2 G_Q$. 
This renormalization law \cite{Efetov} was recently
applied to conductance of granular metals. 
The Eqs.~(\ref{RGEquation}) 
could be also derived in the framework of functional 
RG approach to $\sigma$-model of disordered metal.~\cite{Feigelman}.

We solve the RG Eqs.~(\ref{RGEquation}) 
in  general case to obtain 
\begin{eqnarray}
  T_n^{[k]}(E)
&=&T_n^{[k]}y/\left(1-T_n^{[k]}(1-y)\right), 
\label{Solution}\\
\ln({g_0 E_C}/E) &=& -\frac{1}{2}\sum_{n,\,k}
\ln(1-T_n^{[k]}(1-y))
\end{eqnarray}
The first equation gives the renormalized transmission eigenvalues
at a given value $E$ of the upper cutoff in terms
of variable $y(E)$, $0\leq y \leq 1$. 
The second equation implicitly expresses $y(E)$.

\begin{figure}[t]
\includegraphics[width=3.5in]{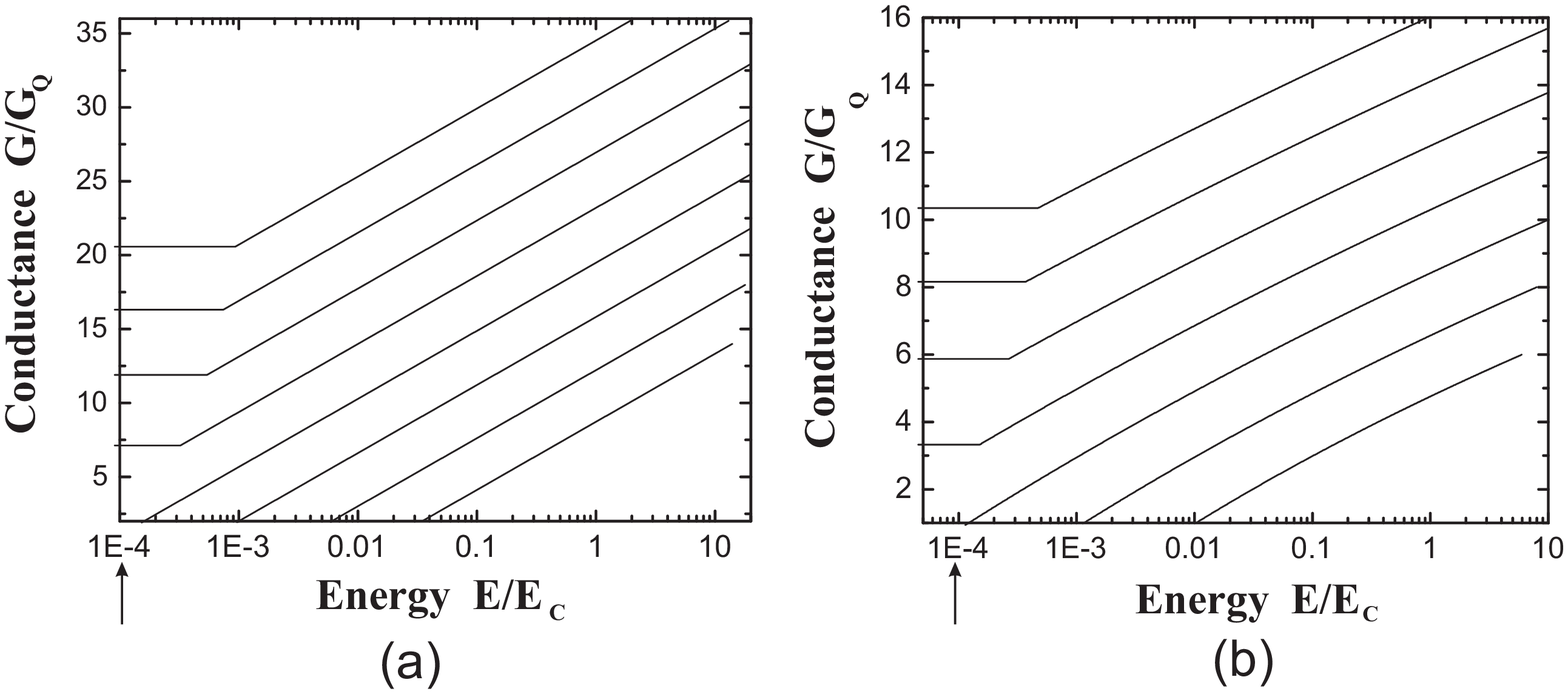}
\caption{
The total conductance of the Coulomb island 
versus the energy: two scenarios. 
We assume $\ln(E_c/\delta)=10.0$. Arrows show the energy scale $\sim \delta$.
Pane (a): tunnel connectors, $g_0$ changes from $42$ (upper curve)
to $14$ (lowermost curve) with the step $4$.
Pane (b): diffusive connectors, $g_0$ changes from $18$ to $6$ with the step $2$.
} 
\end{figure}

We note that the energy dependence of transmission coefficients
induced by interaction is very weak provided $G(E) \gg G_Q$:
If energy is changed by a factor of two, 
the conductance is changed by $\sim G_Q$. 
To use the equations 
for evaluation of FCS at given voltages $V^{[k]}$ of the leads,
one takes $T_n^{[k]}(E)$ 
at upper cutoff  $E = {\rm max}_k(V^{[k]})$, and further disregards
their energy dependence.
Then one can follow the lines of Ref.~\cite{NazBag}:
It is convenient to introduce the 
function $S^{[k]}(x) = -\sum_n\ln[1+\frac{1}{2}T_n^{[k]}(x-1)]$
to incorporate all required information about transmission eigenvalues.  
The renormalization of $S^{[k]}$ in terms 
of $y$ is especially simple: 
$S^{[k]}(x,y) = S^{[k]}((x +1)y-1)-S^{[k]}(2y-1)$.
From this one readily finds the conductance of each scatterer, 
$G^{[k]}(y) = 2 G_Q \partial S^{[k]}/\partial x(1,y)$, 
as well as  the renormalized
transmission distribution  
$T^2 \rho^{[k]}(T,y) = 
({2}/{\pi}){\rm Im}  \{\partial S^{[k]}/\partial x\,(1-2/T - i0,y)\}$.

The RG equations (1) have a fixed point at $T^{[k]}_{n}=0,y=0$
that occur at finite energy
\begin{equation}
E=\widetilde E_C = g_0 E_C\prod_{k,n}(1-T^{[k]}_n)^{1/2} 
\end{equation}
This indicates the breakdown of RG and formation of Coulomb blockade
with the exponentially small gap $\widetilde E_C$.
The same energy scale was obtained 
from equilibrium instanton calculation of Ref. \cite{Nazarov}.
For a field theory, one generally expects different physics
and different energy scales for instantons and 
perturbative RG. The fact that these scales are the same
shows a hidden symmetry of the model which is yet to understand.
\cite{understand}

Alternative low-energy behavior is realized if the current cut-off
reaches $E_{\rm Th}= G(E)\delta/G_Q$. (Fig.2) 
The log renormalization of the transmission
eigenvalues stops at this point and their values saturate.
We thus predict a sharp crossover between the two alternative
scenarios, that occur at value of $g_0=g_c$ corresponding to 
$\widetilde E_C \simeq \delta$. This value equals $g_c = \alpha^{-1}\ln(E_C/\delta)$,
where $\alpha = \frac{1}{2}g_0^{-1}\sum_{n,k}\ln(1-T_n^{[k]})$,  
and depends on
transmission distribution of all connectors. 
If all connectors are tunnel juctions, $\alpha_T = 2$. For diffusive
connectors, $\alpha_D = \pi^2/8$ and the energy dependence of the total conductivity is
given by  $g_D(V)\sim g_0\sqrt{\xi}\,{\rm ctg}{\sqrt{\xi}}$, 
$\xi \equiv 2g_0^{-1}\ln(g_0 E_C/eV)$. (Fig. 2)    

\begin{figure}[t]
\begin{center}
\includegraphics[width=3.5in]{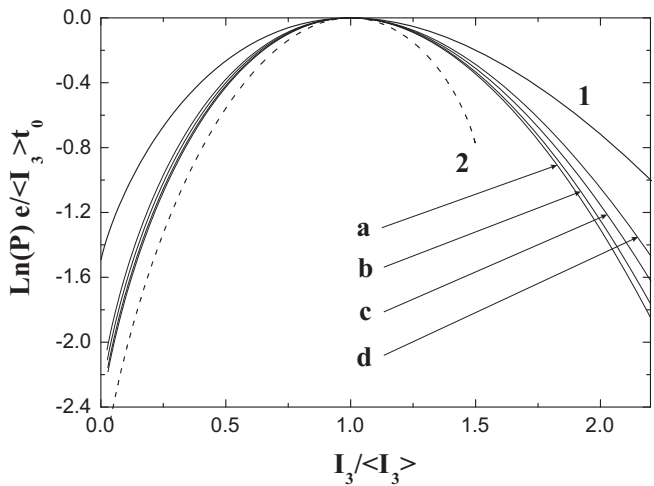}
\caption{An example of FCS: Probability to measure equal currents 
in the 3-terminal 
island with identical junctions. The curves (a-d) present diffusive
connectors at
 different values of the renormalization parameter $y(eV)$:
(a) -- 1.0, (b) -- 0.5, (c) -- 0.1, (d) -- 0.
No interaction effect is seen for tunnel (1) and ballistic (2)
connectors.}
\end{center}
\end{figure}

To give an example of 
FCS calculation in the regime of
weak Coulomb interaction, we consider 
the  3-terminal Coulomb island with identical junctions, 
discussed in \cite{NazBag} (Fig. 1). 
The voltages applied are $V_3=V, \, V_{1,2}=0$
We plot in Fig. 3 the $\log(P)$, 
the logarithm of the probability to measure the same currents 
to the terminals $1$ and $2$, $I_1=I_2=-I_3/2$,
versus the current $I_3$ measured in the terminal $3$ (Fig. 3).
Both $\log(P)$ and $I_3$ are normalized by the average current $<I_3>$,
so that in the absence of interaction 
the curves corresponding to different voltages are the same 
(assuming $eV \gg T$), the shape of the curve reflecting
the transmission distribution in the contacts. To take
interaction into account, we change the transmission eigenvalues
according to Eqs.~(6,7) and evaluate the probability 
with the method of Ref.\cite{NazBag}. 

The curves $1$ (tunnel junctions) and $2$ (ballistic contacts)
stay the same not depending on the renormalization.
Indeed, the renormalization does not affect ballistic transmission,
as to tunnel junctions, it affects only their conductances.
The interaction effect is visible for diffusive junctions.
The curves (a)-(d) correspond to different values of $y(E=eV)$.
The transmission distribution of each contact 
evolves from the diffusive form ($\rho_{D}(T) = g(y)/2T\sqrt{1-T}$)
at the highest voltage ($y \approx 1$)
to the double junction form ($\rho_{DJ}(T) = g(y)/\pi T^{3/2}\sqrt{1-T}$)
at the lowest voltage ($y\approx 0$)~\cite{Kindermann}.  
This visibly changes the form of $\log P(I_3)$.

To conclude, we have analyzed the effect of weak Coulomb interaction 
($G\gg G_Q$) on FCS in Coulomb island. We have revealed that the interaction
effect can be incorporated into a simple energy-dependent renormalization
of transmission eigenvalues, this enables easy evaluation of
all transport properties. The Coulomb blockade develops only if
the total conductance of the island is below some critical
value $\sim G_Q \log(E_C/\delta)$, otherwise the interaction
correction to transport saturates at low energies.

This work is a part of the research program of the "Stichting voor
Fundamenteel Onderzoek der Materie"~(FOM), and we acknowledge the financial
support from the "Nederlandse Organisatie voor Wetenschappelijk Onderzoek"~(NWO).


\begin{thebibliography}{99}
\bibitem{understand} These scales
are different in Ref. \cite{GolubevZaikin}, 
possibly due to the oversimplified
RG they use. The scales were found to be the same in 
Ref. \cite{Feigelman}.
\end{thebibliography}
\end{document}